\documentclass[%showpacs,
showkeys,12pt,
preprint,preprintnumbers,nofootinbib,
groupedaddress,superscriptaddress,amsmath,amssymb]{revtex4}
\usepackage{graphicx}% Include figure files
\usepackage{dcolumn}% Align table columns on decimal point
\usepackage{bm}% bold math
\usepackage{amssymb}
\usepackage{amsmath}
\usepackage{epsfig}    
\usepackage{color}
\usepackage{slashed}
\usepackage{hhline}
\def\be{\begin{equation}}
\def\ee{\end{equation}}
\newcommand{\bea}{\begin{eqnarray}}
\newcommand{\eea}{\end{eqnarray}}
\newcommand{\nn}{\nonumber}

\numberwithin{equation}{section}

\begin{document}

{\begin{flushright}{KIAS-P17044}
\end{flushright}}

%%%%%%%%%
\title{A model with isospin doublet $U(1)_{D}$ gauge symmetry }
%\preprint{KIAS-P14078}
%

\author{Takaaki Nomura}
\email{nomura@kias.re.kr}
\affiliation{School of Physics, KIAS, Seoul 130-722, Korea}

\author{Hiroshi Okada}
\email{macokada3hiroshi@cts.nthu.edu.tw}
\affiliation{Physics Division, National Center for Theoretical Sciences, Hsinchu, Taiwan 300}

\date{\today}

\begin{abstract}
We  propose a model with an extra isospin doublet  $U(1)_D$ gauge symmetry, in which we introduce several extra fermions with odd parity under a discrete $Z_2$ symmetry in order to cancel the gauge anomalies out.  A remarkable issue is that we impose nonzero $U(1)_D$ charge to the standard model Higgs, and it gives the most stringent constraint to the vacuum expectation value of a scalar field breaking the $U(1)_D$ symmetry that are severer than the LEP bound.  We then explore relic density of a Majorana dark matter candidate without conflict of constraints from lepton flavor violating processes. A global analysis is carried out to search for parameters which can accommodate with observed data. 
\end{abstract}
\maketitle
\newpage

\section{Introduction}
Radiatively induced mass models are one of the promising candidate to include a dark matter (DM) candidate naturally, which connect the standard model (SM) fermions and DM candidates playing a role of particles propagating inside a loop diagram for generating radiative masses. Along this line of idea, there exist a lot of papers, {\it i.e.}, ~\cite{a-zee, Cheng-Li, Pilaftsis:1991ug, Ma:2006km, Gu:2007ug, Sahu:2008aw, Gu:2008zf, AristizabalSierra:2006ri, Bouchand:2012dx, McDonald:2013hsa, Ma:2014cfa, Kajiyama:2013sza, Kanemura:2011vm, Kanemura:2011jj, Kanemura:2011mw, Kanemura:2012rj, Farzan:2012sa, Kumericki:2012bf, Kumericki:2012bh, Ma:2012if, Gil:2012ya, Okada:2012np, Hehn:2012kz, Dev:2012sg, Kajiyama:2012xg, Kanemura:2013qva, Law:2013saa, Baek:2014qwa, Kanemura:2014rpa, Fraser:2014yha, Vicente:2014wga, Baek:2015mna, Merle:2015gea, Restrepo:2015ura, Merle:2015ica, Wang:2015saa, Ahn:2012cg, Ma:2012ez, Hernandez:2013dta, Ma:2014eka, Ma:2014yka, Ma:2015pma, Ma:2013mga, radlepton1, Okada:2014nsa, Brdar:2013iea, Okada:2015kkj, Bonnet:2012kz, Joaquim:2014gba, Davoudiasl:2014pya, Lindner:2014oea, Okada:2014nea, Mambrini:2015sia, Boucenna:2014zba, Ahriche:2016acx, Fraser:2015mhb, Fraser:2015zed, Adhikari:2015woo, Okada:2015vwh, Ibarra:2016dlb, Arbelaez:2016mhg, Ahriche:2016rgf, Lu:2016ucn, Kownacki:2016hpm, Ahriche:2016cio, Ahriche:2016ixu, Ma:2016nnn, Nomura:2016jnl, Hagedorn:2016dze, Antipin:2016awv, Nomura:2016emz, Gu:2016ghu, Guo:2016dzl, Hernandez:2015hrt, Megrelidze:2016fcs, Cheung:2016fjo, Seto:2016pks, Lu:2016dbc, Hessler:2016kwm, Okada:2015bxa, Ko:2017quv, Lee:2017ekw, Antipin:2017wiz, Borah:2017dqx, Chiang:2017tai, Kitabayashi:2017sjz, Das:2017ski, Wang:2017mcy,  Nomura:2017ezy, Nomura:2017emk, Nomura:2017vzp, Cheung:2017lpv} at one-loop level.
%%%%%%%

{In constructing radiatively induced mass models, some symmetries are applied to control relevant interactions such as a discrete $Z_2$ symmetry and global/local $U(1)$ symmetry. It would be interesting to assign exotic local $U(1)$ charge to $SU(2)$ doublet leptons since its spontaneous symmetry breaking should be related to generation of Majorana mass of active neutrinos. We are thus interested in $U(1)_D$ gauge symmetry under which the SM fermions with $SU(2)$ doublet are charged.}

In this paper, we propose a model with extra isospin doublet  $U(1)_D$ gauge symmetry, in which we introduce several exotic fermions with odd parity under a discrete $Z_2$ symmetry,
and the neutrino masses are induced at one-loop level.
%%%
Also we discuss the possibility to explain the muon anomalous magnetic moment, and a DM candidate, where the dominant annihilation channel in case of fermionic DM comes from {resonant point of s-channel via the SM Higgs boson.}
A remarkable issue here is that we impose nonzero $U(1)_D$ charge to the SM Higgs doublet
that naturally leads us to the type-II two Higgs doublet model~\cite{Gunion:1989we} in order to generate the SM fermions for up and down sectors.
%%%
Moreover it gives the most stringent constraint on the vacuum expectation value (VEV) of a SM singlet scalar  $\langle \varphi \rangle \equiv v'/\sqrt{2}$ arising in spontaneous breaking of $U(1)_D$; ${\cal O}(10)$ TeV $\lesssim  v'$. 
Thus a smaller extra gauge coupling is in favor of being the smaller DM mass in order to satisfy the correct relic density of  $\Omega h^2  \simeq 0.12$~\cite{Ade:2013zuv}.
Furthermore we investigate constraints from lepton flavor violations (LFVs) which are induced by interactions associated with exotic particles we introduce.

This paper is organized as follows.
In Sec.~II, we show our model, %to introduce exotic fermions and bosons with some additional symmetries,  
and establish  the quark and lepton sector, and derive the analytical forms of neutrino mass matrix, LFVs, muon anomalous magnetic moment and relic  density of DM, and neutral gauge sector,
and we carry out numerical analyses. 
We conclude and discuss in Sec.~III.
%\newpage

%%%%%%%%%%%%%%%%%%%%%%%%%%%%%%%%%%%%%
%\section{The Model}
%\subsection{Model setup}

% \begin{widetext}
\begin{center} 
\begin{table}%[tbc]
%\begin{tiny}
\begin{tabular}{|c||c|c|c|c|c||c|c|c|c|c|c|}\hline\hline  
&\multicolumn{5}{c||}{Quarks} & \multicolumn{5}{c|}{Leptons} \\\hline
Fermions&~$Q_L^\alpha$~&~$u_R^\alpha$~&~$d_R^\alpha$ ~ & ~$Q'^\alpha_{R}$ ~&~$Q''^\alpha_{L}$~&~$L^\alpha_L$~&~$e_R^\alpha$~
&~$L'^\alpha_{R}$~&~$N'^\alpha_{L}$~&~$L''^\alpha_{L}$
%~&~$N''^\alpha_{R}$
\\\hline 
$SU(3)_C$ & $\bm{3}$  & $\bm{3}$& $\bm{3}$ &
 $\bm{3}$  & $\bm{3}$    & $\bm{1}$   & $\bm{1}$  & $\bm{1} $  & $\bm{1} $ & $\bm{1} $
 %& $\bm{1} $ 
   \\\hline 
 %%%
 $SU(2)_L$ & $\bm{2}$  & $\bm{1}$  & $\bm{1}$ & $\bm{2}$ & $\bm{2}$ &
   $\bm{2}$   & $\bm{1}$  & $\bm{2}$   & $\bm{1}$& $\bm{2}$   
 %& $\bm{1}$  
  \\\hline 
 %%%
$U(1)_Y$ & $\frac16$ & $\frac23$  & $-\frac{1}{3}$ & $\frac16$ &  $\frac16$  & $-\frac12$  & $-1$ &  $-\frac12$  &   $0$ &  $-\frac12$ 
 %&$0$  
 \\\hline
 %%%
 $U(1)_{D}$ & $1$ & $0$  & $0$    & $1$ & $0$ & $1$ & $0$  & $1$  & $0$& $0$   
 %& $0$  
 \\\hline
 %%%
$Z_2$ & $+$ & $+$  & $+$
& $-$ & $-$  & $+$ & $+$ & $-$ & $-$ & $-$ 
%& $-$
 \\\hline
%%%
%$\mathbb{Z}_2$ & $+$   & $-$  & $+$ & $+$& $-$& $-$& $+$ & $+$  \\\hline\hline
\end{tabular}
\caption{Field contents of fermions
and their charge assignments under $SU(2)_L\times U(1)_Y\times U(1)\times Z_2$, where each of the flavor index is defined as $\alpha\equiv 1-3$.}
\label{tab:f}
% \end{tiny}
\end{table}
\end{center}
%\end{widetext}

\begin{table}%[thbp]
{
\centering {\fontsize{10}{12}
\begin{tabular}{|c||c|c|c||c|}\hline\hline
&\multicolumn{3}{c||}{VEV$\neq 0$} & \multicolumn{1}{c|}{Inert } \\\hline
  Bosons  &~ $H_u$ ~&~ $H_d$ ~ &~ $\varphi$     ~ &~ $\chi$~  \\\hline
$SU(2)_L$ & $\bm{2}$& $\bm{2}$  & $\bm{1}$   & $\bm{1}$  \\\hline 
$U(1)_Y$ & $\frac12$ & $\frac12$ & $0$ & $0$   \\\hline
 $U(1)_{D}$  & $-1$& $1$ & $1$ & $0$ \\\hline
$Z_2$   & $+$ & $+$& $+$ & $-$  \\\hline
\end{tabular}%
} 
\caption{Boson sector }
\label{tab:b}
}
\end{table}

\section{ Model construction and resulting phenomenology }
In this section, we construct our model and discuss its phenomenology.
In this model, we introduce chiral-flipped mirror quarks and leptons which have opposite $Z_2$ parity to the SM fermions, where
the other charges are the same as the SM SU(2) doublet quark and lepton as can be seen in Table~\ref{tab:f}; we define $Q'_R\equiv [u'_R,d'_R]^T$ and $L'_R\equiv[N'_R,E'_R]^T$, $Q''_L\equiv [u''_L,d''_L]^T$ and $L''_L\equiv[N''_L,E''_L]^T$.
Here we impose an additional $U(1)_D$ gauge symmetry for isospin doublet fields where gauge anomalies associated with $U(1)_D$ cancel between $Q_L(L_L)$ and $Q'_R(L'_L)$.
%%%
On the other hand, double primed exotic fermions $Q''_L$ and $L''_L$ are not charged under the $U(1)_D$ and they are required to give heavy masses of the exotic fermions. 
%Especially, the exotic quarks are severely constrained by the current experimental data  via diphoton processes at LHC, and it suggests that  the typical scale of masses are greater than 500 GeV.
%%%
In the neutrino sector, we introduce left-handed Majorana fermions $N'^\alpha_L$ to generate the masses at one-loop level.
Field contents and their assignments are summarized in Table~\ref{tab:f}, in which $\alpha=1-3$ represents the number of  family.
%%%
%Notice here that we require the third generation couple to the SM-like Higgs directly to maintain consistency of the SM Higgs properties observed by the LHC experiments such as gluon fusion production cross section and branching fractions.
%%%
Under these assignments, the $U(1)_{D}$ gauge symmetry is anomaly free for each generation.
Therefore
\[ [U(1)_Y]^2U(1)_{D}, \ [U(1)_{D}]^2U(1)_Y, \ [U(1)_{D}]^3, \  U(1)_{D}, \] anomalies are zero canceling between the SM fermions and additional fermions.
%%%%%%%%% %%%%%%%%% %%%%%%%%%

As for the scalar sector, we introduce two $SU(2)_L$ singlets $\varphi$ and $\chi$ and {two $SU(2)_L$ doublets $H_u$ and $H_d$,
 where only the $H_{u(d)}$ and $\varphi$ have the VEVs, symbolized by $\langle H_{u(d)}\rangle\equiv v_{u(d)}/\sqrt2$} and $\langle\varphi\rangle\equiv v'/\sqrt2$, breaking the electroweak and $U(1)_D$ symmetries  spontaneously.  On the other hand, we suppose that $\chi$ does not have VEVs that are assured by the $Z_2$ symmetry.
Field contents and their assignments are summarized in Table~\ref{tab:b}, where $\chi$ has to be complex to generate the nonzero masses of the SM neutrinos at one-loop level.
%%%%%%%%% %%%%%%%%% %%%%%%%%%

%{\it Yukawa Lagrangian}:
Under these fields and symmetries, the renormalizable Lagrangians of quark and charged-lepton sector are symbolically found to be 
\begin{align}
-{\cal L}&=
{y^u_{\alpha\beta} \bar Q_{L_\alpha} \tilde H_u u_{R_\beta} + y^d_{\alpha\beta} \bar Q_{L_\alpha}  H_d d_{R_\beta}
+ y^\ell_{\alpha\beta} \bar L_{L_\alpha}  H_d e_{R_\beta}+ y^n_{\alpha\beta} \bar L'_{R_\alpha} \tilde H_u N'_{L_\beta}}
+ M_{N'_\alpha} \bar N'^C_{L_\alpha} N'_{L_\alpha}
%%%
\nn\\&
+f^Q_{\gamma}\bar Q''_{L_\gamma} Q'_{R_\gamma} \varphi^* 
+f^\ell_{\gamma}\bar L''_{L_\gamma} L'_{R_\gamma} \varphi^* 
%%%\nn\\&
+g^Q_{(*)\alpha\beta}\bar Q_{L_\alpha} Q'_{R_\beta} \chi^{(*)} 
+g^\ell_{(*)\alpha\beta}\bar L_{L_\alpha} L'_{R_\beta} \chi^{(*)} 
%%%
+{\rm c.c.}, \label{eq:lag-ykw}
\end{align}
where $f^{Q(\ell)}$ is diagonal without loss of generality,%%%
%$(f''_1)_R\equiv (u''_R, N''_R)$, 
 $(\alpha,\beta,\gamma)=1-3$ are the flavor indices, and $\tilde H_u\equiv \sigma_2 H_u^*$ is the Pauli matrix.

After the electroweak symmetry breaking, these three sectors have their masses of {$m_u\equiv y^u_{\alpha\beta} v_u/\sqrt2$,
$m_d\equiv y^d_{\alpha\beta} v_d/\sqrt2$, and $m_\ell\equiv y^\ell_{\alpha\beta} v_d/\sqrt2$}.

%{\it Higgs potential}:
{Higgs potential is given by
\begin{align}
V&=
\mu_\varphi^2 |\varphi|^2 + \mu_H^2 |H|^2 +\mu^2_{\chi} |\chi|^2  +\mu'^2_{\chi} \left[\chi^2 + \chi^{*2}  \right]
 %+\mu \left[\chi^2 \varphi^* + \chi^{*2} \varphi \right]
+\lambda_\varphi|\varphi|^4 +\lambda_\chi |\chi|^4+ \lambda_{H_u} |H_u|^4+ \lambda_{H_d} |H_d|^4\nn
\\&
+ \lambda_{H_u H_d} |H_u|^2|H_d|^2 + \lambda'_{H_u H_d} |H_u^\dag H_d|^2
+\lambda_{\varphi \chi} |\varphi|^2 |\chi|^2+\lambda_{\varphi H_u} |\varphi|^2|H_u|^2+\lambda_{\varphi H_d} |\varphi|^2|H_d|^2\nn
\\&
+\lambda_{\chi H_u} |\chi|^2|H_u|^2+\lambda_{\chi H_d} |\chi|^2|H_d|^2 {+ \lambda_{H_u H_d \varphi} (H_d^\dagger H_u \varphi \varphi +c.c)} ,
\label{eq:lag-pot}
\end{align}
where the scalar fields are parameterized as 
\begin{align}
%\begin{tiny}
&H_{u,d} =\left[
\begin{array}{c}
w^+_{u,d}\\
\frac{v_{u,d}+h_{u,d}+iz_{u,d}}{\sqrt2}
\end{array}\right],\quad 
%%%
\varphi=\frac{v'+\varphi_R+iz'}{\sqrt2},\quad
%%%
\chi=\frac{\chi_R+i\chi_I}{\sqrt2},
\label{component}
%\end{tiny}
\end{align}
where the {massless states for the mass eigenstates from linear combinations} of $w_{u,d}^\pm$, and $z_{u,d}$ are respectively absorbed by the longitudinal degrees of freedom of charged SM gauge boson $W^\pm$ and neutral SM gauge boson $Z$, and $z'$ is also eaten by neutral $U(1)_D$  gauged boson $Z'$.
{Note that the last term in the scalar potential Eq.~(\ref{eq:lag-pot}) provide quadratic term $H_d^\dagger H_u + c.c$ after $\varphi$ developing the VEV, which allows us to avoid massless Goldstone boson from Higgs doublets.}
As a result, the same amount of bosons are induced from  the type-II two Higgs doublet model;
a physical singly-charged boson($H^\pm$), CP-odd boson($A$) and two CP-even neutral bosons($h,H$),
where $h$ is expected to be the SM Higgs. Notice here that both the mixings between $\varphi_R$ and $(h,H)$ are supposed to be tiny {to avoid the constraints from LHC experiments for Higgs production cross section and branching fraction measurements.} }

\if0
After the spontaneous symmetry breaking, only the CP-even neutral bosons mix each other as follows:
\begin{align}
%%%%%%%
\left[
\begin{array}{c}
h\\ \varphi_R
\end{array}\right]
={\cal O}_ \alpha
\left[
\begin{array}{c}
h_1\\ h_2
\end{array}\right],\quad
{\cal O}_ \alpha
\equiv
\left[
\begin{array}{cc}
-c_\alpha & s_\alpha\\ 
s_\alpha & c_\alpha \\
\end{array}\right],
\end{align}
where we define $c_\alpha\equiv \cos\alpha$, $s_\alpha\equiv \sin\alpha$,
 $h_i(i=1,2)$ is the mass eigenstate of the neutral boson with VEVs. Here $h_1$ is the SM-like Higgs and $h_2$ is the additional Higgs boson.
 All of the mass eigenvalues and mixings are written in terms of VEVs, and quartic couplings in the Higgs potential after inserting the tadpole conditions: $\partial V/\partial h|_{v,v'}=0$ and $\partial V/\partial \varphi_R|_{v,v'}=0$. 
% \textcolor{red}{If you like, please write each of values explicitly...}
Here we omit the detailed analysis of the scalar sector and assume that the mixing is sufficiently small to avoid the constraints from LHC measurements of SM Higgs production cross section and branching fractions.
\fi

% {\it Charged exotic fermions}:
\subsection{Fermion masses}

{\it  Exotic neutral fermions}:
The charged exotic fermions are mass eigenstates after the $U(1)_D$ spontaneous breaking, that is,
$M'^Q\equiv f^Q v'/\sqrt2$ and $M'^\ell\equiv f^\ell v'/\sqrt2$ which can have heavy mass, as TeV scale or larger, due to the VEV of $\varphi$.

%let us focus on the neutral sector below.
%%% 
Here we discuss the neutral fermion sector in the following.
We have a mass matrix of neutral fermion in basis of  $\Psi\equiv [N'^C_R,N'_L,N''_L]^T$,
and they are given by
\begin{align}
M_{\Psi}
&\equiv
\left[\begin{array}{ccc}
{\bf 0}_{3\times3} & { (m_n)}_{3\times3} &  (M'^\ell)_{3\times3} \\ 
{ (m_n^T)}_{3\times3} & {\bf (M_{N'})}_{3\times3}&0 \\ 
  (M'^{\ell T})_{3\times3} & 0 & 0\\
\end{array}\right],
 \end{align}
where ${ (m_n)}_{\alpha\beta} \equiv (y^n)_{\alpha\beta}v_u/\sqrt2$.
Then the mass eigenstate and its mixing is respectively defined by $D_{\psi}=V M_{\Psi} V^T$, and 
\begin{align} 
\left[\begin{array}{c}
N'^C_R  \\ 
N'_L \\ 
N''_L \\ 
\end{array}\right]_i
=(V^T)_{ij} \psi_j , \  { i,j=1\sim9},
\end{align}
where $V$ is  the  unitary mixing matrix with six by six, and $\psi_i$ is the mass eigenstate, and $D_{\psi}$ is mass eigenvalue.
%%%

%%%%%%%%%%%%%%%%%%%
\begin{figure}[t]
\begin{center}
\includegraphics[width=70mm]{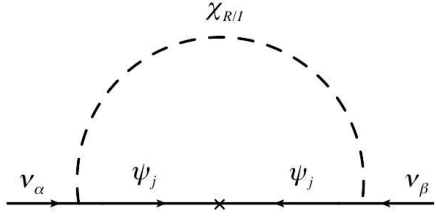} 
\caption{The one loop diagram to generate neutrino masses. } 
  \label{fig:neut}
\end{center}\end{figure}
%%%%%%%%%%%%%%%%%%%

{\it Active neutrinos}:
The active neutrino mass matrix is induced at the one-loop level in fig.~\ref{fig:neut}; the analytic form
 is given by
\begin{align}
&(m_\nu)_{\alpha\beta}\approx 
-\frac{1}{(4\pi)^2}\sum_{a,b=1}^3\sum_{j=1}^9
(g^{\ell})_{\alpha a}V^\dag_{a+2,j} D_{\psi_j} (g^\ell)_{\beta b} V^\dag_{b+2,j} F_I^j(r_{R_j},r_{I_j}),\\
%%%
&F_I(r_1,r_2)=\frac{r_1 \ln[r_1]-r_2\ln[r_2]+r_1r_2\ln[r_2/r_1]}{(1-r_1)(1-r_2)},
\end{align}
where $r_{R_j(I_j)}\equiv[m_{\chi_{R(I)}}/D_{\psi_j}]^2$.
%%%
Since one diagonalizes neutrino mass matrix as $D_\nu\approx V_{MNS}m_\nu V_{MNS}^T$, 
we can rewrite Yukawa coupling in terms of neutrino oscillation data and some parameters as~\cite{Casas:2001sr}:
\begin{align}
(g^\ell)_{3\times3} = (V^\dag_{MNS} \sqrt{D_\nu} O R^{-1/2}  V)_{3\times3},
\end{align}
where $ R _\alpha  \equiv D_{\psi_\alpha}F_I^\alpha(r_{R_\alpha},r_{I_\alpha})$, and $O$ is an arbitral three by nine orthogonal matrix: $OO^T=1_{3\times3}$.
Satisfying the neutrino oscillation data is rather easy task due to $O$, and all we should take care is the constraints of lepton flavor violations via $g^\ell$. 
%%%%%%%%%

\subsection{Muon $g-2$ and LFVs}

 {\it Muon $g-2$}:
The muon anomalous magnetic moment($\Delta a_\mu$) has been observed and its discrepancy from the SM is estimated by~\cite{Hagiwara:2011af}
\begin{align}
\Delta a_\mu=(26.1\pm8.0)\times10^{-10}.\label{eq:exp-g2mu}
\end{align}
%%%
Our $\Delta a_\mu$ is induced at one-loop level via the Yukawa interactions associated with $g^\ell$ where the $\chi_{R(I)}$ and $E'$ propagate inside the loop diagram.
The analytic form is computed as
\begin{align}
\Delta a_{\mu}&\approx \frac{2m_\mu^2}{(4\pi)^2} \sum_{\gamma=1-3}
g^\ell_{2\gamma}(g^{\ell\dag})_{\gamma2} F_{II}(m_\chi,M'^\ell_{\gamma}),\\
%%%
F_{II}(m_a,m_b)&\equiv\frac{2 m_a^6+3m_a^4m_b^2-6m_a^2m_b^4+m_b^6+12m_a^4m_b^2\ln\left[\frac{m_b}{m_a}\right]}{12(m_a^2-m_b^2)^4},
\end{align}
where $m_\chi\equiv m_{\chi_{R}}\approx m_{\chi_{I}}$, and $M'^\ell_{\gamma}$ is the mass of charged extra fermions.
%%%

{\it Lepton flavor violations (LFVs)}: LFV processes of $\ell \to \ell' \gamma$ are arisen from the same term as the $(g-2)_\mu$, and their forms are given by
\begin{align}
BR(\ell_\alpha\to \ell_\beta \gamma)
&\approx\frac{48\pi^3C_{ab} \alpha_{em}}{(4\pi)^4G_F^2}
\left| \sum_{\gamma=1-3}
g^\ell_{\beta\gamma}(g^{\ell\dag})_{\gamma\alpha} F_{II}(m_\chi,M'^\ell_{\gamma}) \right|^2,
\end{align}
where $\alpha_{em}\approx1/137$ is the fine-structure constant, $G_F\approx1.17\times10^{-5}$ GeV$^{-2}$ is the Fermi constant,
and $C_{21}\approx1$, $C_{31}\approx 0.1784$, $C_{32}\approx0.1736$. 
Experimental upper bounds are given by~\cite{TheMEG:2016wtm, Adam:2013mnn}: 
\[{\rm BR}(\mu\to e \gamma)\lesssim 4.2\times 10^{-13},\ 
{\rm BR}(\tau\to e \gamma)\lesssim 3.3\times 10^{-8},\ 
{\rm BR}(\tau\to \mu \gamma)\lesssim 4.4\times 10^{-13},
\]
where we define $\ell_1\equiv e$,  $\ell_2\equiv \mu$, and  $\ell_3\equiv \tau$. 

\subsection{$Z'$ neutral vector boson}
{\it $Z_{SM}$-$Z'$ mixing}:
Since $H$ has nonzero $U(1)_D$ charge, there is mixing between $Z_{SM}$ and $Z'$, where $Z'$ is the extra gauge boson via $U(1)_D$. 
The resulting mass matrix in basis of $(Z_{SM},Z')$  is given by
\begin{align}
m_{Z_{SM}Z'}^2
&= \frac14
\left[\begin{array}{cc}
(g_1^2+g_2^2) v^2 &  -2\sqrt{g_1^2+g_2^2}g' v^2  \\ 
-2 \sqrt{g_1^2+g_2^2}g' v^2  & 4 g'^2 (v^2+ v'^2)   \\ 
\end{array}\right]\nn\\
&=
%%%
m_{Z'}^2
\left[\begin{array}{cc}
\epsilon_1^2 & -\epsilon_1 \epsilon_2  \\ 
-\epsilon_1 \epsilon_2 & 1+\epsilon_2^2  \\ 
\end{array}\right],
\end{align}  
where $m_{Z_{SM}}\equiv \frac{\sqrt{g_1^2+g_2^2}v}{2}\approx 91.18$ GeV, $m_{Z'}\equiv  g'v'$, $\epsilon_1\equiv \frac{m_{Z_{SM}}}{m_{Z'}}$, $\epsilon_2\equiv \frac{v}{v'}$, $g_1$, $g_2$, and $g'$ are gauge coupling of $U(1)_Y$, $SU(2)_L$, and $U(1)_D$, respectively.
Then its mass matrix is diagonalized by the two by two mixing matrix $V$ as $V m_{Z_{SM}Z'}^2 V^T%=D_Z^2
\equiv {\rm Diag}(m^2_{Z_{}},m^2_{Z_{D}}) $,
where we work under $\epsilon_2^2<<1$ and
\begin{align}
m^2_{Z}&\approx m_{Z_{SM}}^2(1-\epsilon_2^2),\
m^2_{Z_{R}}\approx m_{Z'}^2 (1 +  \epsilon_1^2\epsilon_2^2),\label{eq:zm}
\\
%%%
V&\approx
\left[\begin{array}{cc}
c_{Z} &  s_{Z} \\ 
-s_{Z}  &  c_{Z}  \\ 
\end{array}\right], \quad \theta_{Z} = \frac{1}{2} \tan^{-1} \left[ \frac{2 \epsilon_1 \epsilon_2}{1+\epsilon_2^2-\epsilon_1^2} \right].
\end{align} 
Since the ambiguity of the $Z$ boson mass is around $0.0021$~\cite{Olive:2016xmw}:
\begin{align}
|\Delta m_Z|=m_{Z_{SM}}\left(\sqrt{1-\epsilon_2^2}-1\right)\lesssim 0.0021\ {\rm GeV},
\end{align}       
one finds the stringent constraint on the $v'$ from Eq.(\ref{eq:zm}) as
\begin{align}
v' \gtrsim 36.25\ {\rm TeV}.
\end{align}    
%%%%%%%%%%%%%%%%%%%%

\if0
%%%%%%%%%%%%%%%%%%%
\begin{figure}[t]
\begin{center}
\includegraphics[width=100mm]{mx-gp.eps} \qquad
\caption{}
  \label{fig:mx-gp}
\end{center}\end{figure}
%%%%%%%%%%%%%%%%%%
\fi

  {\it Constraint from LEP experiment } :
  Since our $Z'$ universally couples to SM leptons the LEP experiment provides the strongest constraints on the gauge coupling and $Z'$ mass.
  Assuming $m_{Z'} \gtrsim 200$ GeV, the LEP constraint is applied to the effective Lagrangian 
\begin{equation}
L_{eff} = \frac{1}{1+\delta_{e \ell}} \frac{g'^2}{m_{Z'}^2} (\bar e \gamma^\mu {P_L }e)( \bar \ell \gamma_\mu {P_L } \ell) \label{eq:eff}
\end{equation}
where $\ell = e$, $\mu$ and  $\tau$.
We then obtain following constraint from the analysis of data by measurement at LEP~\cite{Schael:2013ita}: 
%which  tells us the following restriction:
\begin{align}
\frac{m_{Z'}}{g'} \gtrsim  {4.0}\ {\rm TeV}.\label{eq:lep}
\end{align}
This bound is weaker than the constraint of the $Z_{SM}$-$Z'$ mixing. 
%%%
{Then we finally find the relation as
follows:
\begin{align}
\frac{m_{Z'}}{g'} %\approx \frac{2 M_X}{g'} 
\approx  \sqrt{v^2+v'^2}\gtrsim 36.251 \ {\rm TeV}.\label{eq:const}
\end{align}
where we have used $v\approx 246$ GeV.
}
%%%
%In fig.~\ref{fig:mx-gp}, we show a plot of $M_X/g'$ in terms of $M_X$, where the horizontal line represents the experimental bound given by Eq.~(\ref{eq:lep}) and the region below the line is excluded. Each of  the black, blue, and red line is the one for $g'=0.01$, $g'=0.05$, and $g'=0.2$. One finds that $g'=0.01$ allows all the region of the DM mass, while $g'=0.05$ and $g'=0.2$ allow the DM mass up to 1 GeV and 1200 GeV respectively, combining the figure 1.

\begin{figure}[t]
\begin{center}
\includegraphics[width=100mm]{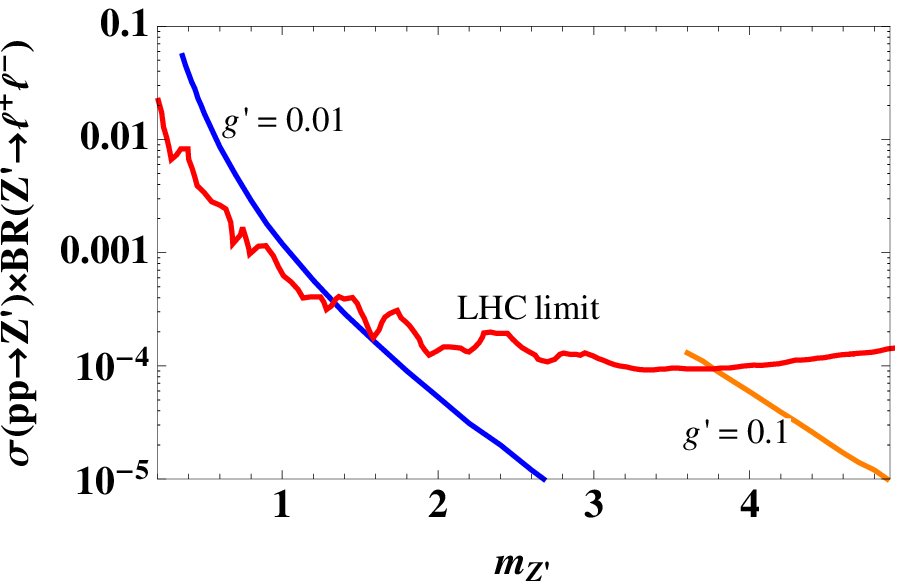} \qquad
\caption{$\sigma(pp \to Z') BR(Z' \to \ell^+ \ell^-)$ as a function of $m_{Z'}$ with $\sqrt{s}=13$ TeV where red curve indicate the upper limit by the ATLAS experiment~\cite{ATLAS:2017wce}. The blue and orange lines correspond to $g'=0.01$ and $g'=0.1$ respectively which start from the region satisfying Eq.~(\ref{eq:const}).} 
  \label{fig:ZpLHC}
\end{center}\end{figure}

{
{\it Z' production at the LHC } :
Our $Z'$ boson can be produced at the LHC since it couples to SM quarks, and the most significant signature is obtained from the process $pp \to Z' \to \ell^+ \ell^-$ ($\ell = e, \mu$).
Here we estimate the cross section at the LHC 13 TeV with {\it CalcHEP} 3.6~\cite{Belyaev:2012qa} implementing relevant interactions and using the CTEQ6 parton distribution functions (PDFs)
~\cite{Nadolsky:2008zw}.  Fig.~\ref{fig:ZpLHC} shows $\sigma(pp \to Z') BR(Z' \to \ell^+ \ell^-)$ as a function of $m_{Z'}$ for $g'=0.01$ and $g'=0.1$, which is compared with the latest upper limit given by the ATLAS experiment~\cite{ATLAS:2017wce}; note that the lines start from lower bound of $m_{Z'}$ from Eq.~(\ref{eq:const}). We find that the LHC limit further excludes the parameter space for small $g'$(lighter $Z'$) region, and the constraint from $Z_{SM}$-$Z'$ mixing becomes stronger for larger $g'$ region. Further parameter space will be tested with more integrated luminosity in future LHC experiments.
}

%%%%%%%%%%%%%%%%%%%
%%%%%%%%%%%%%%%%%%

{
\subsection{Dark matter}
In case of boson DM candidate $\chi$, 
we have several annihilation channels via $g^\ell$ Yukawa term and potential term.
As we have shown some figures in the global analysis in Fig.~2, however, the typical order of $g^\ell$ are $0.01$ to satisfy all the constraints such as LFVs, and it gives ${\cal O}(10^{-12}\sim10^{-11})$ scale of muon $g-2$~\cite{Chiang:2017tai}. Therefore the Yukawa coupling cannot be dominant to satisfy the correct relic density $\Omega h^2\approx0.12$~\cite{Ade:2013zuv}.
Also we have to satisfy the constraint of the direct detection experiment such as LUX~\cite{Akerib:2016vxi} via Higgs portals.
The typical order to evade this bound is to take the correspond quartic parameter $\lambda_{\chi H_u,\chi H_d,\varphi\chi }\lesssim {\cal O}(0.01)$~\cite{Das:2017ski} in case of the SM Higgs portal.
Even with these kinds of constraints, one could find wide allowed region to satisfy the correct relic density without conflict of our allowed space of global analysis~in Fig.~2.
%%% To satisfy the correct relic density $\Omega h^2\approx0.12$~\cite{Ade:2013zuv} in bosonic case, the typical order of Yukawa couplings are $0.01$ that at most gives ${\cal O}(10^{-12})$ scale of muon $g-2$~\cite{Chiang:2017tai}.

Here we will focus on and  and analyze the lightest  Majorana DM candidate $\psi_1\equiv X$.
In case of fermion DM, 
we also have several annihilation channels via $g^\ell, y^n,f^\ell$ Yukawa terms and kinetic term.
But $g^\ell$ cannot be dominant due to the same reason of bosonic DM case.\\
%%%
In the kinetic term, one finds that  its cross section is proportional to the form $\left| \frac{g'}{m_{Z'}}\right|^4 M_X^2\lesssim 5.8\times10^{-19}\frac{M_X^2}{\rm GeV^4}$ from Eq.~(\ref{eq:const}). Since the DM mass is at most 1 TeV from the global analysis, the cross section is found to be less than of the order $10^{-12}$\ GeV$^{-2}$, which is much lower than the typical cross section to satisfy the observed relic density of  DM $\langle \sigma v \rangle \sim10^{-9}$ GeV$^{-2}$. Considering the chirality suppression, the kinetic mode via $Z'$ cannot be dominant
even if a resonant point at $M_X\sim m_{Z'}/2$.~\footnote{We have checked that relic density does not reach 0.12 for the whole range of DM mass $1\sim 1200$ GeV that is allowed by the global analysis in Table~\ref{fig:mx-lfvs}.}\\
%%%
In case of relying on the resonance point at the half mass of the CP-even Higgses.
We could find solutions at the half of masses except the SM Higgs resonance~\cite{Kanemura:2010sh}. 
Instead of this trivial solution, we discuss the quark interactions via Yukawa couplings $g^Q$. 
The thermally averaged cross section is d-wave dominant in terms of $v_{\rm rel}$ expansion approximation, and its form is  given by 
\begin{align}
\sigma v_{\rm rel}(2X\to Q_\alpha\bar Q_\gamma)\approx \frac{M_X^6}{40\pi}\left|\frac{(g^Q)_{\alpha \beta}(g^Q)^\dag_{\beta \gamma}}{(M_X^2+M^2_{Q'_\beta})^2}\right|^2 v^4_{\rm rel},
\end{align}
where we assume to be $g^Q\approx g^Q_*$, and $(\alpha,\beta,\gamma)$ are implicitly summed over 1-3.
The resulting relic density is found to be
\begin{align}
&\Omega h^2\approx 
\frac{3.57\times 10^7 x_f^3}{\sqrt{g_*(x_f)} M_{\rm PL} d_{\rm eff}},   \label{eq:relic-rl}
\end{align}
respectively, where the present relic density is $0.1199 \pm 0.0054$ at the 2$\sigma$ confidential level (CL)~\cite{Ade:2013zuv}, $g_*(x_f\approx 25)\approx100$ counts the degrees of freedom for relativistic 
particles, and $M_{\rm PL}\approx 1.22\times 10^{19}$~GeV is the Planck mass.
\footnote{One might induce the semi-leptonic rare decays through box types of diagrams in $g^Q$ that give several constraints on $g^Q$.
However these contributions identically vanishes when the real scalar runs inside the loop. Thus one does not need to worry about these kinds of constraints even in the case of large $g^Q$.}
%%%

 \subsection{Global analysis}
First of all we focus on the analysis of DM, since it does not depend on the other phenomenologies except the mass of DM. 
 Then we randomly select the input parameters as
\begin{align}
g^Q_{11,22,33}\in (0.1,\sqrt{4\pi}),\quad M_X\in (1,5000)\ {\rm GeV},\quad M_{Q'_{1,2,3}}\in (1.2 M_X,6000)\ {\rm GeV},
\end{align}
where we have simplified the mass matrix $g^Q$ to be diagonal, and the upper value of $\sqrt{4\pi}$ comes from the perturbative limit, and the lower bound on $M_{Q'_{1,2,3}}^{\rm Min}= 1.2 M_X$ is taken to evade the coannihilation processes for simplicity.
%%%
Fig.~\ref{fig:mx-mQp} represents $M_X$ and $M_{Q'_1}$, which shows the allowed mass range of DM is at around $10$ GeV$\lesssim M_X\lesssim$ 1200 GeV, where the Yukawa couplings runs whole
the range that we have taken, and we have adopted the relaxed range of relic density; $0.11\le \Omega h^2\le 0.13$ instead of tight value 0.12.

%{\it Global analysis}:
Next we have a global analysis taking into account the neutrino oscillation data, constraints from LFVs, muon $g-2$. 
Then we also randomly choose the same input parameter ranges as the ones of DM analysis.
Figs.~\ref{fig:mx-lfvs} represent the muon $g-2$ in terms of $M_X$, where the green region shows the allowed region in the DM analysis.
%%% %%%
It shows that the maximal muon $g-2$ is ${\cal O}(10^{-11})$, which is smaller than the expected value by three order of magnitudes.

%%%%%%%%%%%%%%%%%%%
\begin{figure}[t]
\begin{center}
\includegraphics[width=70mm]{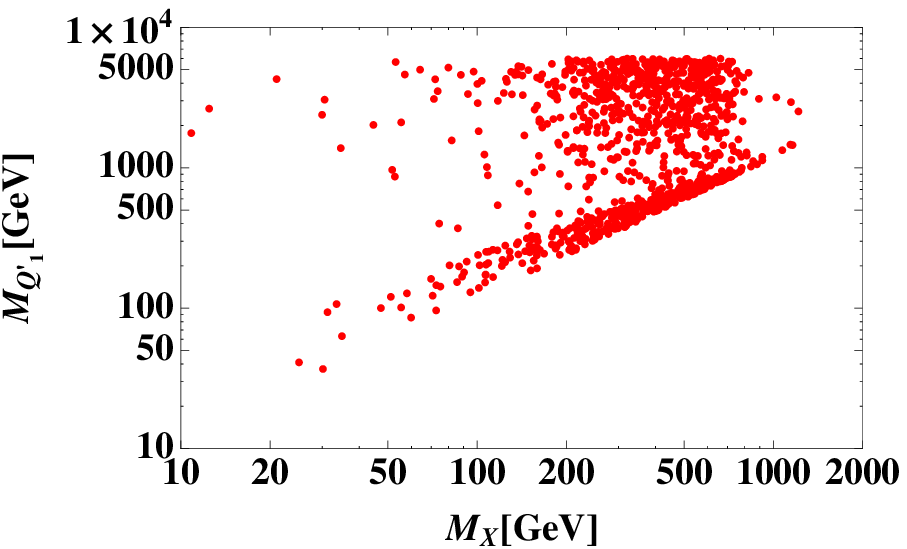} \caption{The  figure represents $M_X$ and $M_{Q'_1}$.} 
  \label{fig:mx-mQp}
\end{center}\end{figure}
%%%%%%%%%%%%%%%%%%%
%%%%%%%%%%%%%%%%%%%
\begin{figure}[t]
\begin{center}
\includegraphics[width=70mm]{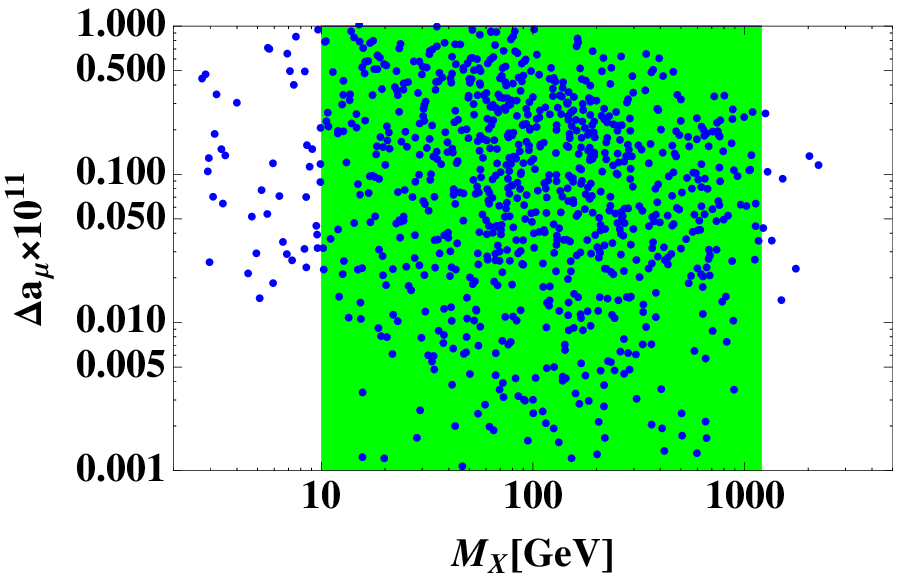} 
\caption{The  figure represents the muon $g-2$ in terms of $M_X$, where the green region shows the allowed region in the DM analysis.} 
  \label{fig:mx-lfvs}
\end{center}\end{figure}
%%%%%%%%%%%%%%%%%%%

\section{ Conclusions and discussions}
We have proposed a model with extra isospin doublet  $U(1)_D$ gauge symmetry, in which we introduce several exotic fermions with odd $Z_2$ parity,
and the neutrino masses are induced at one-loop level. Also we have discussed the possibility to explain the muon anomalous magnetic moment,
 and a Majorana dark matter candidate without conflict of constraints of LFVs.
%, where the dominant channel in case of fermionic DM comes from the Yukawa term.
In our global analysis with wide ranges, we have found that the maximal value of muon $g-2$ is of the order $10^{-11}$ as can be seen in the upper plots in fig.~\ref{fig:mx-lfvs}, which is smaller than the experimentally expected value by three order magnitude.
The allowed mass region is found to be $10$ GeV$\le M_X\le$1200 GeV.
One of the remarkable issue is that we  have imposed nonzero $U(1)_D$ charge to the standard model (SM) Higgs, and it gives the most stringent constraint to the vacuum expectation value $v'$ arising from spontaneous symmetry breaking of $U(1)_D$.
Therefore we have obtained 36.25 TeV $\lesssim v'$; this constraint is even stronger than the LEP constraint for $m_{Z'}/g' \simeq v'$. 
{Thus the kinetic contribution to the relic density cannot be dominant, when considering fermionic DM. 
And the dominant annihilation channel comes from resonant point of s-channel via the SM Higgs boson.}
We have also discussed $Z'$ production at the LHC and find some parameter region is excluded at the current data where constraints from $v'$ becomes stronger for larger $g'$ region. Further parameter region will be explored by the future LHC data.

%\section*{ Appendix}
%%%%%%%%%%%%%%%%%%%...

%\newpage
%%%%%%%%%%%%%%%%%%%%%%%%%%%%%%%%%%%
%\hspace{0.2cm} {\bf Acknowledgments}
%\section*{Acknowledgments}:
%\vspace{0.5cm}
\section*{Acknowledgments}
\vspace{0.5cm}
H. O. is sincerely grateful for all the KIAS members, Korean cordial persons, foods, culture, weather, and all the other things.
%%%%%%%%%%%%%%%%%%%%%%%%%%%%%%%%%%%
%%%%%%%%%%%%%%%%%%%%%%%%%%%%%%%%%%%

\if0
\section*{Appendix}
One finds the squared amplitude for the process $X\bar X\to \nu_a\bar\nu_b$ from the Yukawa term of $g^\ell$ as follows:
\begin{align}
|\bar {\cal M}(X\bar X\to\nu_a\bar \nu_b)|^2 & \approx 
4\left| \sum_{\beta\beta'=1}^3 
\frac{(g^\ell)_{a\beta}V^\dag_{\beta1} V_{1\beta'} (g^{\ell\dag})_{\beta'b}}{t-m_\chi^2}\right|^2(p_1\cdot k_1)(p_2\cdot k_2)\nn\\
&+4\left|\sum_{\beta\beta'=1}^3 \frac{(g^\ell)_{a\beta}V^\dag_{\beta1} V_{1\beta'} (g^{\ell\dag})_{\beta'b}}{u-m_\chi^2}\right|^2(p_1\cdot k_2)(p_2\cdot k_1)\nn\\
%
&
-\frac12 \sum_{\beta\beta'=1}^3 \left(
\frac{(g^\ell)_{a\beta}V^\dag_{\beta1} }{t-m_{\chi}^2} \frac{V_{1\beta'} (g^{\ell\dag})_{\beta'b}}{u-m_{\chi}^2}
+
\frac{(g^\ell)_{a\beta}V^\dag_{\beta1} }{u-m_{\chi}^2} \frac{V_{1\beta'} (g^{\ell\dag})_{\beta'b}}{t-m_{\chi}^2}
\right) M_X^2 (k_1\cdot k_2).%
\end{align}
%and each of inner products such as $(p_1\cdot p_2)$ is found to be ref.~\cite{Cheung:2016ypw}.
\fi

\end{document}